\documentclass[reprint,aps,pra]{revtex4-1}

\usepackage{bm}

\begin{document}

\title{Sommerfeld Fine-Structure  Formula for Two-Body Atoms}

\author{John H. Connell}
\email{connell@stcc.edu}
\affiliation{Springfield Technical Community College\\
Springfield, Massachusetts, USA}


\date{\today}

\begin{abstract}

For relativistic atomic two-body systems such as the hydrogen atom, positronium, 
and muon-proton bound states, a two-body generalisation of the single-particle 
Sommerfeld fine-structure formula for the relativistic bound-state energies is 
found.  The two-body Sommerfeld bound-state energy formula is obtained from a 
two-body wave equation which is physically correct to order $
(Z\alpha )^4$.  
The two-body  Sommerfeld formula makes two predictions in order $
(Z\alpha )^6$ for every 
bound state and every mass ratio. With $N$ the Bohr quantum number:  (a) The 
coefficient of the $(Z\alpha )^6/N^6$ energy term has a specified value which depends only 
on the masses of the bound particles, not on angular quantum numbers; (b) The 
coefficient of the $(Z\alpha )^6/N^4$ energy term is a specified multiple of the {\em square\/} of 
the coefficient of the $(Z\alpha )^4/N^3$ energy term.  Both these predictions are verified 
in positronium by previous calculations to order $(Z\alpha )^
6$ which used second-order 
perturbation theory.  They are also correct in the Coulomb-Dirac limit.  The 
effect of the two-body Sommerfeld formula on calculations of muon-proton 
bound-state energies is examined.

\end{abstract}

\maketitle

\section{INTRODUCTION}

The lack of an analytically solvable relativistic wave equation for two-body 
atomic systems has compelled physicists to use second-order perturbation theory 
in calculating energy levels to order $(Z\alpha )^6$ in systems such as positronium 
\cite{Jacek,Cz}.  Several two-body wave equations or prescriptions (reviewed in 
ref.  \cite{1991}), starting with the Breit equation \cite{Breit} in 1929, have 
correctly described the energy levels of two-body atomic systems to order $
(Z\alpha )^4$ 
using first-order perturbation theory.  But none has been solvable analytically, so 
none has been able to be a starting-point for the calculation of atomic energy 
levels at order $(Z\alpha )^6$ without the use of second-order perturbation theory.  

For the rest of this paper the symbol $\alpha$ will be a synonym  for $
Z\alpha$.  No radiative 
corrections are considered.

To explain what is found in this paper and what remains to be found, we remind 
the reader that the analytic solution \cite{1928} to the Coulomb Dirac equation in 
1928 gave the one-particle Sommerfeld fine-structure formula for the bound-state 
energies:  
\begin{equation}E_{\text{Dirac}}=\frac m{\sqrt {1+\frac {\alpha^
2}{(n+\epsilon_{\text{Dirac}}+1)^2}}}.\label{E0}\end{equation}
Here $m$ is the mass of the electron, $\alpha$ is the fine-structure constant, $
n$ is the 
radial quantum number and $\epsilon_{\text{Dirac}}$ is determined by the eigenvalue of a 
finite-dimensional angular equation 
\begin{equation}(\epsilon_{\text{Dirac}}+1)^2\xi =[(1+\bm{\sigma}
\cdot\bm{L})^2-\alpha^2]\xi .\label{e0}\end{equation}
Because equation (\ref{e0}) for the angular parameter $\epsilon_{\text{
Dirac}}$ is exactly 
solvable, the bound-state energies predicted by equation (\ref{E0}) can be evaluated 
by simple algebra not only to order $\alpha^4$, but to order $
\alpha^6$ and even $\alpha^8$.  The 
Coulomb-Dirac problem does not need second- and third-order perturbation theory.

If a two-body wave equation accurate to order $(Z\alpha )^4$ were to be 
solvable analytically like the one-particle Coulomb Dirac equation, no second-order 
perturbation theory would be needed to obtain two-body energy levels to order 
$(Z\alpha )^6$.  The other terms needed to complete the physical calculation to order 
$(Z\alpha )^6$ would only need to be evaluated in first-order perturbation theory.

\section{Main Result}

In this paper we will find a two-particle counterpart to the Sommerfeld 
fine-structure formula (\ref{E0}).  It is 
\begin{equation}E=\sqrt {m^2+M^2+\frac {2mM}{\sqrt {1+\frac {
\alpha^2}{(n+\epsilon +1)^2}}}}\label{SF}\end{equation}
Here $M$ is the mass of the other particle.  It is easy to see that $
E-M$ reduces 
to the one-particle Sommerfeld fine-structure formula (\ref{E0}) when $
M\rightarrow\infty$.  The 
formula (\ref{SF}) for the bound-state energies is obtained from a two-body 
relativistic atomic wave equation which was derived in ref.  \cite{1991} and is 
repeated below.

In this work we will not be able to derive an angular equation for the angular 
parameter $\epsilon$ which is a two-particle counterpart to equation (\ref{e0}).  Therefore 
we have only half the analytic solution.

Nevertheless, from the two-particle Sommerfeld formula (\ref{SF}) alone, assuming 
that $\epsilon$ exists without knowing it, we have found two predictions for order $
\alpha^6$ 
bound-state energy terms for any masses, which are verified by previous calculations for 
positronium which had to use second-order perturbation theory.  These results 
raise the hope that it may eventually be possible to find a full analytic solution 
to a relativistic atomic two-body wave equation which would eliminate the need 
for second-order perturbation calculations to obtain energies to order $
\alpha^6$.

\section{Two-Body Wave Equation}

The local relativistic atomic two-body wave equation (\ref{we}) below was derived 
in ref.  \cite{1991} from the Bethe-Salpeter equation for two spin-$
1/2$ constituent 
point particles bound by a single-photon-exchange kernel in the Coulomb gauge.  
The derivation used a simple quasi-potential approximation \cite{1967,1971} with its 
associated Blankenbecler-Sugar correction series \cite{1966}.

The masses of the bound particles are denoted by $m,M$. The  bound-state 
energy $E$ is parametrised by a quantity $\beta$ as follows:
\begin{equation}E=\sqrt {m^2-\beta^2}+\sqrt {M^2-\beta^2}\label{E}\end{equation}
The particles' individual bound-state energies also occur:
\begin{equation}t=\sqrt {m^2-\beta^2}\mbox{\rm ,}\qquad T=\sqrt {
M^2-\beta^2}\label{tT}\end{equation}
The Pauli matrices $\bm{\sigma}$ and the Dirac matrices $\bm{
\gamma}$, $\gamma^0$ refer to the 
particle of mass $m$, while $\bm{\Sigma}$, $\bm{\Gamma}$ and $
\Gamma^0$ refer to the particle of mass 
$M$.  The operator ${\bf p}$ is $-i\bm{\nabla}$, where $\bm{
\nabla}$ refers to the relative 
position co\"ordinate ${\bf r}$.  Also $r=|{\bf r}|$ and $\hat {
{\bf r}}\equiv {\bf r}/r$.  

The relativistic bound-state wave equation in the centre-of-mass system is
\begin{eqnarray}
&&[{\bf p}^2+\beta^2]\psi ({\bf r})\nonumber\\*
&&=-\frac 1{2E}\left[m-\bm{\gamma}\cdot {\bf p}+\gamma^0t\right
]\left[M+\bm{\Gamma}\cdot {\bf p}+\Gamma^0T\right]\times\nonumber\\*
&&\times\left\{-\gamma^0\Gamma^0\frac {\alpha}r+\frac {\bm{\gamma}
\cdot\bm{\Gamma}+\bm{\gamma}\cdot\hat {{\bf r}}\bm{\Gamma}\cdot
\hat {{\bf r}}}2\frac {\alpha}r\right.+\label{we}\\*
&&+\left.\frac {1+\gamma^0}2\frac {1+\Gamma^0}2\frac 1{2E}\left
(\frac {\alpha}r\right)^2-\frac {g-2}{4M}\bm{\gamma}\cdot\hat 
{\bf r}\times\bm{\Sigma}\frac {\alpha}{r^2}\right\}\psi ({\bf r}
)\nonumber\end{eqnarray}
The eigenvalue is $\beta^2$, not $E$.  The constant $\beta^2$ is substituted into the square roots 
in equation (\ref{E}) to obtain the energy $E$.  There are no non-local operators of 
the form $\sqrt {{\bf p}^2+m^2}$, so no kinetic-energy correction terms $
{\bf p}^4$, ${\bf p}^6$ appear in 
perturbation theory.  

The first and second terms in the curly brackets are the standard binding 
potential and Breit interaction.  The third term is an effective term equal to 
$1/2E$ times the square of the binding potential.  This is a consequence of the 
Blankenbecler-Sugar correction formalism, which ensures that the relativistic 
energies are correct \cite{SLAC}.  The final term in the curly brackets, which 
was not included in ref.  \cite{1991}, contains an anomalous magnetic moment for 
the particle of mass $M$.  It is derived by adding a term $i
(g-2)({\bf k}\times\bm{\Sigma})/4M$ to 
the Dirac matrix $\bm{\Gamma}$ in the vertex function of the particle of mass $
M$ in the 
originating Bethe-Salpeter equation.  The term is included here so that the 
particle of mass $M$ may represent a proton.

As discussed in ref.  \cite{1991}, the wave equation (\ref{we}) is expected to be 
valid for $r\agt\alpha /2E$, because the Blankenbecler-Sugar correction formalism results 
in a series in powers of $(-\alpha /r)/2E$.  In addition, for the hydrogen atom and 
muonic hydrogen \cite{Ant, Pohl}, $\alpha /2E$ is about $0.0
01$ fm.  The mean-square 
charge radius of the proton is about $1$ fm.  Therefore except perhaps for 
positronium the validity of the two-body wave equation (\ref{we}) is expected to 
end well before $r$ descends to $\alpha /2E$.

In the limit $M\rightarrow\infty$, dividing through by the first right-hand operator in square 
brackets, it is easy to see that equation (\ref{we}) reduces to the one-body 
Coulomb Dirac equation 
\begin{equation}\left[m+\bm{\gamma}\cdot {\bf p}-\gamma^0t\right
]\psi ({\bf r})=\frac {1+\Gamma^0}2\gamma^0\frac {\alpha}r\psi 
({\bf r})\label{we0}\end{equation}
which gives the energies (\ref{E0}) and the angular equation (\ref{e0}).

To verify the correctness of the wave equation (\ref{we}) we give its bound-state 
energies to order $\alpha^4$. With $J$ (often written $F$) denoting the total spin of the 
bound state, notation used is:
\begin{equation}\mu =\frac {mM}{m+M}\qquad x=\frac {\mu^2}{m
M}\qquad\rho =\frac {\mu}M-\frac {\mu}m\label{muxrho}\end{equation}
\begin{equation}a^2=\left[\rho +(g-2)x\right]^2L(L+1)\label{asq}\end{equation}
From first-order perturbation theory with $N$ the Bohr quantum number the 
bound-state energies of the wave equation (\ref{we}) are 
\begin{eqnarray}
E&=&m+M-\frac {\alpha^2\mu}{2N^2}+\alpha^4\mu\frac {3-x}{8N^
4}-\nonumber\\
&-&\frac {\alpha^4\mu}{2N^3}\left[(1-\delta_{L0})\Theta +\delta_{
L0}\Phi\right]\label{E4}\end{eqnarray}
in which
\begin{equation}\Theta =\left\{\begin{array}{ll}
\frac 1L+\frac {2xg}{(2L+1)(2J+1)}\quad&L=J+1\\
\frac 1{L+1}-\frac {2xg}{(2L+1)(2J+1)}\quad&L=J-1\\
\frac 12\left[\frac 1L+\frac 1{L+1}+\frac {\sqrt {1+4a^2}}{(2L+1)
L(L+1)}\right]\quad&L=J,S\approx 1\\
\frac 12\left[\frac 1L+\frac 1{L+1}-\frac {\sqrt {1+4a^2}}{(2L+1)
L(L+1)}\right]\quad&L=J,S\approx 0\end{array}
\right.\label{theta}\end{equation}
and
\begin{equation}\Phi =\left\{\begin{array}{ll}
1-2xg/3&\quad L=0,S=1\\
1+2xg&\quad L=0,S=0\end{array}
\right.\label{phi}\end{equation}
The symbols $S\approx 1$, $S\approx 0$ for $L=J$ stand for the state in which $
S$ is 
predominantly $1$ or $0$, respectively.  

Examining the results above, we find that in the limit $M\rightarrow
\infty$ the energies of the 
Dirac Coulomb equation are recovered.  For $m\ll M$ it is easy to verify that the 
hyperfine splittings are correct to order $m/M$ (\cite{BS}, Sec.  22).  With $
m=M$ 
and $g=2$ we find the standard positronium energies without the annihilation term 
(\cite{BS}, Sec.  23).  These results verify that to order $
\alpha^4$, the two-body wave 
equation (\ref{we}) gives physically correct results.

\section{Derivation of the  Two-Body  Energy Formula}

Following conventional treatments of the Coulomb Schr\"odinger equation and the 
Coulomb Dirac equation, we substitute 
\begin{equation}\psi ({\bf r})=e^{-\beta r}r^{\epsilon}\sum_
ja_jr^j\label{wfn}\end{equation}
into the wave equation (\ref{we}).  Here the coefficients $a_
j$ are 16-dimensional 
vectors.  The expansion is expected to be valid for $r\agt\alpha 
/2E$, or for $r$ greater 
than the proton radius if a proton is present.   One obtains a four-term 
recurrence relation for the coefficients $a_j$.  

When the dominant terms acting on the large component of the wave function in 
the wave equation (\ref{we}) are examined, we see that they are the same as in 
the Coulomb Schr\"odinger equation.  That means that if the series (\ref{wfn}) does 
not terminate for some $j=n$, the wavefunction will diverge as $
e^{+\beta r}$ for large $r$.  
So that the wavefunction (\ref{wfn}) will go to zero at infinity, in this paper we 
will {\em assume\/} that the series terminates at $j=n$.  Then it is easy to show that 
with $a_j=0$ for $j>n$, the recurrence relation implies this equation for $
a_n$:  
\begin{eqnarray}
&&2\beta (\epsilon +1+n)a_n\nonumber\\
&=&-\frac {\alpha}{2E}\tilde {m}\tilde {M}\left[-\gamma^0\Gamma^
0+\frac {\bm{\gamma}\cdot\bm{\Gamma}+\bm{\gamma}\cdot\hat {{\bf r}}\bm{
\Gamma}\cdot\hat {{\bf r}}}2\right]a_n\label{an}\end{eqnarray}
containing the abbreviations
\begin{equation}\tilde {m}=m-i\beta\bm{\gamma}\cdot\hat {{\bf r}}
+\gamma^0t,\qquad\tilde {M}=M+i\beta\bm{\Gamma}\cdot\hat {{\bf r}}
+\Gamma^0T.\label{mM}\end{equation}

Equation (\ref{an}) shows that $a_n=\tilde {m}\tilde {M}b.$ Substituting that back into (\ref{an}) 
puts a projection operator $\tilde {m}\tilde {M}$ on each side of the Coulomb and Breit terms; 
multiplying them out gives  a scalar multiple of $\tilde {m}
\tilde {M}$ again.  One thus finds 
\[2\beta (\epsilon +n+1)a_n=-\frac {\alpha}{2E}\left[-4tT+4\beta^
2\right]a_n\nonumber\]
which is to say
\begin{equation}\frac {\beta E}{tT-\beta^2}=\frac {\alpha}{\epsilon 
+n+1}\label{bE}\end{equation}
Using equations (\ref{E}) and (\ref{tT}), equation (\ref{bE}) gives the two-body 
Sommerfeld ``fine-structure'' bound-state energy formula (\ref{SF}).  

The parameter $\epsilon$, which depends on the angular quantum numbers, cannot be found 
from the two-body wave equation (\ref{we}) as easily as $\epsilon_{\text{
Dirac}}$ is found 
from the Coulomb Dirac equation $(\ref{we0})$.  The reason is that $
\epsilon_{\text{Dirac}}$ is 
determined by the regular singularity of (\ref{we0}) as $r\rightarrow 
0$.  But for finite $m$ and 
$M$, the wave equation (\ref{we}) is expected to be valid only for $
r$ exceeding the 
small but finite value $\alpha /2E$, or exceeding the radius of a proton if a proton is 
one of the bound particles.   While the two-body Sommerfeld energy formula 
(\ref{SF}) has come from the requirement that the wavefunction remain finite at 
large $r$, we cannot use the behaviour of eqn.  (\ref{we}) as $
r\rightarrow 0$ to determine $\epsilon$.  

However, the assumption that the quantity $\epsilon$ does exist will lead below to some 
new predictions which are verified in known physical systems such as the 
hydrogen atom and positronium.  The predictions apply to muon-proton bound 
states as well.

\section{Expansion of E in  Powers of $\alpha^2$}

In the bound-state energy formula (\ref{SF}) the quantity $n
+\epsilon +1$ occurs.  While 
in the Coulomb Schr\"odinger equation $\epsilon$ is always $
L$, in the Coulomb Dirac case 
equation (\ref{e0}) sometimes gives $\epsilon_{\text{Dirac}}
\approx L$, but sometimes 
$\epsilon_{\text{Dirac}}\approx L-1$.  In the latter case the ground state is known to be given by 
$n=1,$ not $n=0$.  Thus  in either case, $n+\epsilon_{\text{
Dirac}}+1$ is still approximately 
the Bohr quantum number $N$.  In fact it is easy to find from (\ref{e0}) that,
with $\kappa =|{\bf L}+\bm{\sigma}/2| +1/2$,
\begin{equation}n+\epsilon_{\text{Dirac}}+1=N-\frac {\alpha^
2}{2\kappa}-\frac {\alpha^4}{8\kappa^3}-\frac {\alpha^6}{16\kappa^
5}+\cdots\label{ne10}\end{equation}
It is likely that the same thing happens for the two-body 
wave equation (\ref{we}). Therefore we expand 
\begin{equation}n+\epsilon +1=N+\alpha^2\epsilon_2+\alpha^4\epsilon_
4+\alpha^6\epsilon_6+\cdots\label{ne1}\end{equation}

When the expansion  (\ref{ne1}) is substituted into the two-body Sommerfeld formula 
(\ref{SF}), the energy levels $E$ can be expanded in powers of $
\alpha$$^2$.  Recalling that 
$x\equiv \mu^2/mM$, one finds 
\begin{equation}E=m+M+\alpha^2\mu C_2+\alpha^4\mu C_4+\alpha^
6\mu C_6+\alpha^8\mu C_8+\cdots\label{Eexpn}\end{equation}
in which
\begin{eqnarray}
C_2&=&-\,\frac 1{2N^2}\label{C2}\\
C_4&=&\frac {3-x}{8N^4}+\frac {\epsilon_2}{N^3}\label{C4}\\
C_6&=&-\,\frac {5-3x+x^2}{16N^6}-\frac {(3-x)\epsilon_2}{2N^
5}-\frac {3\epsilon_2^2}{2N^4}+\frac {\epsilon_4}{N^3}\label{C6}\\
C_8&=&\frac {35-29x+18x^2-5x^3}{128N^8}+\nonumber\\*
&+&\frac {3\left[5-3x+x^2\right]\epsilon_2}{8N^7}+\frac {5(3
-x)\epsilon_2^2}{4N^6}+\label{C8}\\*
&+&\frac {4\epsilon_2^3-(3-x)\epsilon_4}{2N^5}-\frac {3\epsilon_
2\epsilon_4}{N^4}+\frac {\epsilon_6}{N^3}\nonumber\end{eqnarray}

In the Coulomb-Dirac limit, $\mu =m$ and $x=0$.  In this limit, using the expansion of 
$\epsilon_{\text{Dirac}}$ from eqn.  (\ref{ne10}), eqns.  (\ref{Eexpn}) to (\ref{C8}) reproduce 
the known expansion of $E_{\text{Dirac}}$ to order $\alpha^8$.  For general masses $
m$,$M$ we 
now examine the expansion (\ref{Eexpn}) and the coefficients (\ref{C2}) to 
(\ref{C8}) to see whether first, they agree with the known results to order $
\alpha^4$, 
and second, to see whether they predict anything not known yet.

\section{$\alpha^2$ and $\alpha^4$ terms}

The coefficient $C_2$ gives the Bohr levels correctly. 

Next, referring to the energy levels (\ref{E4}) given above, even the coefficient 
$C_4$ suggests the correctness of the two-body Sommerfeld formula (\ref{SF}).  Eqn.  
(\ref{C4}) agrees that there are $\alpha^4/N^4$ and $\alpha^
4/N^3$ terms only.  Moreover, as 
eqn.  (\ref{E4}) shows, $C_4$ gives the $\alpha^4/N^4$ term correctly for all masses, 
verifying that it is independent of the unknown value of $\epsilon$.  Finally, from the 
first-order perturbation theory result (\ref{E4}) we can read off the value of $
\epsilon_2$:  
\begin{equation}\epsilon_2=-\frac 12\left[(1-\delta_{L0})\Theta 
+\delta_{L0}\Phi\right]\label{e2}\end{equation}
with $\Theta$ and $\Phi$ given by   (\ref{theta}) and   (\ref{phi}).

\section{Predictions of $\alpha^6$ Terms}

The $\alpha^6$ terms in (\ref{C6}) will tell whether the two-body Sommerfeld energy 
formula (\ref{SF}) contains any information previously  obtainable only from 
using second-order perturbation theory, and whether it predicts anything as yet 
uncalculated. Note that the predictions described hold automatically in the 
Coulomb-Dirac limit $M\to\infty$.

\subsection{$\alpha^6/N^6$ Terms}

The coefficient of the $\alpha^6\mu /N^6$ term given by (\ref{C6}) does not need knowledge 
of the angular eigenvalue $\epsilon$.  It is predicted to have the same value 
$-[5-3x+x^2]/16$ for every state, depending only on the masses.  Therefore for 
positronium, for which $\mu =m/2$ and $x=1/4$, eqn.  (\ref{C6}) predicts that the 
coefficient of $\alpha^6m/N^6$ is always $-69/512$.  In fact, precisely this value has been 
found for all P-states \cite{P}, for all S-states \cite{Cz}, and in Zatorski's recent 
calculations for all $L\ge 2$ states \cite{Jacek}.  To contrast the methods, we note 
that in ref.  \cite{Jacek} the coefficient $-69/512$ is the sum of five terms, three 
from second-order perturbation theory (eqns.  (159), (160) and (162) of ref.  
\cite{Jacek}) and two first-order terms, one of which contains the expectation 
value of the ${\bf p}^6/m^5$ kinetic-energy correction which is absent in our formalism 
(ref.  \cite{Jacek} eqns.  (91) and (122)).

In addition, Pachucki \cite{KP} has found the same function $
-[5-3x+x^2]/16$ for the 
coefficient of the $\alpha^6\mu /N^6$ energy term as in (\ref{C6}) for arbitrary masses --- 
but only for S-states.  Accordingly this coefficient remains to be tested for $
L\ge 1$ 
when $m\ne M$.

\subsection{$\alpha^6/N^4$ Terms}

For $L\ge 1$, in ref.  \cite{Jacek} the $\alpha^6m/N^4$ terms of the positronium energy 
levels come from the sum of six second-order perturbation terms (eqns.  (153), 
(158), (164), (171), (174) and (177)) and no first-order terms.  Since the two-body 
Sommerfeld formula (\ref{SF}) is a candidate to replace the results of 
second-order perturbation theory once $\epsilon$ is fully determined, these calculated 
$\alpha^6/N^4$ terms are a possible second test of the Sommerfeld formula (\ref{SF}).  

In positronium, from (\ref{C4}) the $\alpha^4/N^3$ energy term is $
(\epsilon_2/2)\alpha^4m/N^3$, and 
from (\ref{C6}) the $\alpha^6/N^4$ energy term is $(-3\epsilon_
2^2/4)\alpha^6m/N$$^4$.  Therefore it is 
predicted that the coefficient of the $\alpha^6m/N^4$ term of positronium should be $
-3$ 
times the \textit{square} of the coefficient of the $\alpha^
4m/N^3$ term.  

As an example for $L=J+1$, with $m=M$ and $g=2$, from (\ref{E4}) and (\ref{theta})
above the coefficient of $\alpha^4m/N^3$ in positronium is
\[\frac {-(4L^2+L-1)}{4L(2L-1)(2L+1)}.\]
In eqn.  (211) of ref.  \cite{Jacek} the calculated coefficient of the $
\alpha^6m/N^4$ term 
for $L=J+1$ is given from the sum of the six second-order terms as 
\[-\,\frac {3-6L-21L^2+24L^3+48L^4}{16L^2(2L-1)^2(2L+1)^2}\]
It can be seen that the latter expression is indeed $-3$ times the square of the 
former.  This verifies the prediction of the two-body Sommerfeld formula for 
positronium $L=J+1$ states.  The prediction also gives the other $
\alpha^6m/N^4$ terms 
of positronium energies correctly (eqns.  (207), (215) and (219) of ref.  
\cite{Jacek}).  

For the $^1S_0$ and $^3S_1$ states of positronium the coefficients of the $
\alpha^6m/N^4$ term 
contain other contributions (see ref. \cite{Cz}) and no prediction can be made. For 
$L\ge 1$ and $m\ne M$  the prediction for the $\alpha^6/N^4$ term remains to be tested.

\subsection{$\alpha^6/N^5$  and $\alpha^6/N^3$ Terms}

While (\ref{C6}) predicts the coefficient of $\alpha^6\mu /N^
5$ to be $-(3-x)\epsilon_2/2$, in which $\epsilon_2$ 
is already known from (\ref{e2}), we cannot tell whether this coefficient is 
confirmed by previous positronium calculations because many terms of order $
\alpha^6$ 
contain $1/N^5$ \cite{Jacek,Cz}.  

The final coefficent $\epsilon_4$ of $\alpha^6\mu /N^3$ in (\ref{C6}) cannot be evaluated and tested 
until an angular equation for $\epsilon$ analogous to (\ref{e0}) is found from which $
\epsilon_4$ can 
be calculated.  Note, however, that since $L=J\pm 1$ mixing is prevented in order $
\alpha^4$ 
because $\langle L=J-1|1/r^3|L=J+1\rangle =0$, this mixing does not occur in $
\epsilon_2$ either, 
because of eqn.  (\ref{e2}).  Consequently, such mixing can only occur in the $
\epsilon_4$ 
term.  Therefore eqn.  (\ref{C6}) predicts that in order $\alpha^
6$,  $L=J\pm 1$ mixing 
should occur in the $\alpha^6/N^3$ term only.

\section{Other Remarks}

\subsection{$ $$m\ll M$}

For $m\ll M$  and $\epsilon =\epsilon_{\text{Dirac}}$  the two-body Sommerfeld formula reproduces a 
result in the review of hydrogen-like atoms by Eides, Grotch and Shelyuto
\cite{GESH}. The two-body Sommerfeld formula   (\ref{SF}) can be written, once 
again recalling that $x=\mu /(m+M)$, as
\begin{equation}E=(m+M)\sqrt {1+2xh}\label{Ex}\end{equation}
with
\begin{equation}h=\frac 1{\sqrt {1+\frac {\alpha^2}{(n+\epsilon 
+1)^2}}}-1\label{h}\end{equation}
For $m\ll M$, $x$ is small.  Equation (\ref{Ex}) expanded in powers of $
x$ up to order 
$x^2$ is 
\begin{equation}E=(m+M)+\mu h-\frac 12\frac {\mu^2}{m+M}h^2.\label{Egesh}\end{equation}
With $\epsilon =\epsilon_{\text{Dirac}}$ this is the same expression that appears in eqn.  (38) of 
ref.  \cite{GESH}.  Expansion of expression (\ref{Egesh}) to order $
\alpha^6$ reproduces 
the expression (\ref{C6}) for $C_6$ except for the $x^2$ term, which is small for the 
hydrogen atom.  This agreement represents another confirmation of the two-body 
Sommerfeld formula (\ref{SF}).

\subsection{$(Z\alpha )^8$}

The expansion of the bound-state energy given by equations (\ref{Eexpn}) to 
(\ref{C8}) is carried out to order $\alpha^8$, meaning $(Z\alpha 
)^8$.  As a single example, in the 
unlikely event that any energy levels of positronium are calculated to order $
(Z\alpha )^8$ 
for some angular quantum numbers and all Bohr quantum numbers $
N$, eqn.  
(\ref{C8}) predicts that the coefficient of $(Z\alpha )^8m/N^
8$ for every state of positronium 
will be $1843/16384$.

\subsection{Evaluation from Bethe-Salpeter Kernel}

The wave equation (\ref{we}) was derived from a Bethe-Salpeter equation whose 
kernel consists of the exchange of a single Coulomb-gauge photon.  To evaluate 
physical energies to order $(Z\alpha )^6$ fully, the procedure would be (i) to evaluate the 
energies of this binding Bethe-Salpeter equation fully to order $
(Z\alpha )^6$, then (ii) to 
evaluate the energy contributions of other relevant kernels, including crossed 
diagrams, all of which would need only first-order expectation values to obtain 
energies to order $(Z\alpha )^6$.  

In part (i) the first task is evidently to find and solve a two-body angular 
equation similar to (\ref{e0}) for the unknown angular parameter $
\epsilon$, so that the 
two-body Sommerfeld formula (\ref{SF}) can be evaluated to any desired accuracy.  
Then the Blankenbecler-Sugar correction formalism \cite{1991} can be used to 
evaluate the remaining corrections.  

Since for real two-body atoms the wave equation (\ref{we}) has not revealed a 
way to calculate $\epsilon$ yet, we report on a complete implementation of part  (i) for a 
far simpler model.  

Ref.  \cite{2001} used a Bethe-Salpeter equation for scalar constituent particles of 
masses $m,M$ bound by a scalar-Coulomb kernel $-4mM\alpha /r$.  Following through the 
formalism of ref.  \cite{1991} gave a two-body wave equation with a solution 
which could be evaluated fully.  The wave equation had a regular singularity as 
$r\rightarrow 0$ which gave an algebraic equation for $\epsilon$ not unlike eqn.  (\ref{e0}) of the 
present paper.  The solution series terminated at $j=n$ for appropriate energies.  
The resultant two-body Sommerfeld formula was quite similar to eqn.  (\ref{SF}) 
of the present paper.  It gave energies to order $\alpha^6$ since $
\epsilon$ was known.  The 
Blankenbecler-Sugar correction formula was used to find the corrections to the 
approximations which led to the wave equation.  Evaluation of their expectation 
values gave $\alpha^6$ terms for all $L$, as well as $\alpha^
5$ and $\alpha^6\log\alpha$ terms for $L=0$.  Thus 
the calculation of the bound-state energies of the model scalar Bethe-Salpeter 
equation was completed to order $(Z\alpha )^6$ without using second-order perturbation 
theory.  

It is possible that a similar implementation for real two-body atoms using a 
Coulomb-gauge kernel would require a return to the derivation in ref.  \cite{1991}, 
perhaps obtaining a wave equation sightly different to (\ref{we}) near the origin.

\subsection{Muonic Hydrogen}

For muon-proton bound states we have compared the energies calculated by 
first-order perturbation theory (\ref{E4}) to an evaluation using the exact 
two-body Sommerfeld formula (\ref{SF}) in which, given the value of $
n+\epsilon +1$, 
recoil effects appear to be exact. 

Substituting the approximation 
\[n+\epsilon +1=N+\alpha^2\epsilon_2\]
into the two-body Sommerfeld formula (\ref{SF}) with $\epsilon_
2$ given by the first-order 
perturbation value (\ref{e2}), it was found that the energy of the nominal 1S${}_{
1/2}^{F=0}$ 
state evaluated by the two-body Sommerfeld formula was $0.00
5$ meV lower than 
when evaluated by first-order perturbation theory.  The other 1S state had a 
smaller difference.  The energy of the nominal 2S${}_{1/2}^{
F=0}$ state was $0.0008$ meV lower 
when calculated by the Sommerfeld formula than by perturbation theory.  The 
other $N=2$ states had a smaller difference.  Thus for muonic hydrogen the recoil 
correction to first-order perturbation theory given by the Sommerfeld formula 
appears to be small for the 2P and 2S states of current interest \cite{Ant,Pohl}.

\section{Conclusion}

Verification of the two $(Z\alpha )^6$ predictions, especially for the $
(Z\alpha )^6/N^4$ term which 
was quite unexpected, indicates two things.  (i) The two-body Sommerfeld energy 
formula (\ref{SF}) seems to be true.  (ii) The angular eigenvalue $
\epsilon$ apparently 
exists.  We do not know yet whether the two-body atomic wave equation 
(\ref{we}), valid as written for $r\agt \alpha /2E$, determines a finite-dimensional angular 
equation for $\epsilon$ analogous to the one-particle angular equation (\ref{e0}) for 
$\epsilon_{\text{Dirac}}$.  

It is expected that any complete analytic solution to the relativistic two-body 
atomic bound-state problem will give a two-body Sommerfeld formula for the 
bound-state energies which would give the $(Z\alpha )^4$ levels completely, and which 
would also give the contribution of order $(Z\alpha )^6$ which previously has had to be 
calculated from second-order perturbation theory on the Breit interaction.  To 
obtain complete physical results to order $(Z\alpha )^6$, as was done for the simple scalar 
Bethe-Salpeter equation in ref.  \cite{2001},   it is necessary first to go back to 
the originating Bethe-Salpeter equation and evaluate its energies to order $
(Z\alpha )^6$.  
This is done by letting the formalism given in \cite{1991} determine what small 
additional terms are present beyond those contained in the Sommerfeld formula.  
The expectation values of these terms are evaluated.  

More than 80 years have passed since the analytic solution to the Coulomb Dirac 
equation gave the Sommerfeld fine-structure formula (\ref{E0}).  The existence of 
the two-body Sommerfeld formula (\ref{SF}), even though its angular eigenvalue $
\epsilon$ 
remains unknown for now, suggests that a wave equation for two-body atoms, 
perhaps somewhat similar to our equation (\ref{we}), may eventually be found and 
solved analytically.


\begin{thebibliography}{99}

\bibitem{Jacek}
J. Zatorski, Phys. Rev. A \textbf{78}, 032103 (2008).
\bibitem{Cz}
A. Czarnecki, K. Melnikov and A. Yelkhovski, Phys. Rev. A  \textbf{59}, 4316 (1999).
\bibitem{1991}
J.H. Connell, Phys. Rev. D \textbf{43}, 1393 (1991).
\bibitem{Breit} G. Breit, Phys. Rev. \textbf{34}, 553 (1929).
\bibitem{1928} W. Gordon, Z. Physik \textbf{48}, 11 (1928); C.G. Darwin, 
Proc. Roy. Soc. Ser. A \textbf{118}, 654 (1928).
\bibitem{1967} J. H. Connell, Thesis, University of Washington (Seattle, 1967).
\bibitem{1971} I.T. Todorov, Phys. Rev. D \textbf{3}, 2351 (1971).
\bibitem{1966} R. Blankenbecler and R. Sugar, Phys. Rev. \textbf{142}, 1051 (1966).
\bibitem{SLAC} J.H. Connell, SLAC-PUB 5633 (1991).
\bibitem{Ant} A. Antognini, F. Kottmann, F. Biraben, P. Indelicato, 
F. Nez and R. Pohl, \url{arxiv.org/abs/1208.2637}.
\bibitem{Pohl} R. Pohl, R. Gilman, G. A. Miller and K. Pachucki, 
Ann. Rev. Nucl. Part. Sci. (2013); \url{arxiv.org/abs/1301.0905}.
\bibitem{BS} H.A. Bethe and E.E. Salpeter, \textit{Quantum Mechanics of
One- and Two-Electron Atoms} (Springer-Verlag/Academic, New York, 1957) [also published
in \textit{Handbuch der Physik}, Vol. XXXV (Springer, Berlin/G\"ottingen/Heidelberg,
1956)].
\bibitem{P} I.B. Khriplovich, A.I. Milstein and A.S. Yelkhovski, Zh. Exp. Teo. Fiz. 
\textbf{105}, 299 (1994). This paper's results are quoted in refs.  \cite{Jacek} and  \cite{Cz}. 
\bibitem{KP} K. Pachucki, Phys. Rev. Letters \textbf{79}, 4120 (1997).
\bibitem{GESH} M. I.  Eides,  H.  Grotch \& V. A.   
Shelyuto, {{\em Theory of Light Hydrogen-like Atoms}}
(Springer Tracts in Modern Physics $\bm{222}$, 2006);
Phys. Rep. \textbf{342}, 63 (2001);  \url{arxiv.org/abs/hep-th/0002158}.
\bibitem{2001} J. H. Connell,   \url{arxiv.org/abs/hep-th/0006082}.  

\end{thebibliography}
\end{document}